\documentclass[a4paper]{article}

\usepackage{INTERSPEECH2022}
\usepackage{multirow}
\usepackage{multicol}
\usepackage{caption}
\usepackage{subcaption}

\title{Membership Inference Attacks Against Self-supervised Speech Models}
\name{Wei-Cheng Tseng, Wei-Tsung Kao, Hung-yi Lee}
\address{
  Graduate Institute of Communication Engineering, National Taiwan University}
\email{r09942094@ntu.edu.tw, r09942067@ntu.edu.tw, hungyilee@ntu.edu.tw}

\begin{document}

\maketitle
\begin{abstract}
Recently, adapting the idea of self-supervised learning (SSL) on continuous speech has started gaining attention. SSL models pre-trained on a huge amount of unlabeled audio can generate general-purpose representations that benefit a wide variety of speech processing tasks. 
Despite their ubiquitous deployment, however, the potential privacy risks of these models have not been well investigated. 
In this paper, we present the first privacy analysis on several SSL speech models using Membership Inference Attacks (MIA) under black-box access.
The experiment results show that these pre-trained models are vulnerable to MIA and prone to membership information leakage with high Area Under the Curve (AUC) in both utterance-level and speaker-level.
Furthermore, we also conduct several ablation studies to understand the factors that contribute to the success of MIA.
\end{abstract}


\section{Introduction}
\label{sec:intro}
As the applications of deep learning become more and more widespread, it is inevitable for people to pay extra attention to the privacy issues of deep learning models. 
Several works have been proposed to inspect the privacy-preserving ability of deep learning models by applying privacy attacks against them.
\cite{fredrikson2015model} finds it possible to reconstruct recognizable images from the facial recognition system by exploiting the output predictions.
\cite{tramer2016stealing} successfully duplicates the functionality from a Machine Learning as a Service (MLaaS) system under black-box access.
\cite{shokri2017membership} take a leaf out of Differential Privacy \cite{dwork2008differential} and investigates the information leakage from the viewpoint of data.

Among the privacy attacks mentioned above, Membership Inference Attack \cite{shokri2017membership} (MIA) focus on the privacy of the individuals whose data was used to train the model. Given a model and an exact datapoint, the adversary infers whether this datapoint was used to train the model or not.
It is considered one of the simplest privacy attack and can serve as a canary of more severe privacy issues \cite{de2020overview}.
A significant amount of research work has been investigated in the context of a wide variety of datasets and machine learning models \cite{chen2020gan, hayes2019logan, song2019auditing, salem2018ml, he2021node, choquette2021label}.
In the speech processing community, the privacy issues of some important applications such as ASR also have been explored~\cite{shah2021evaluating}.
However, it is worth noting that these previous works mostly only focused on supervised learning.

Beyond supervised learning, in recent years, self-super-vised learning (SSL) pre-trained models have become an important component of natural language processing (NLP) and speech processing. 
SSL speech models can be pre-trained on large-scale unlabeled speech datasets by solving discriminative task~\cite{hsu2021hubert, baevski2020wav2vec, oord2018representation}, generative task~\cite{liu2020tera,liu2020mockingjay} or in multi-task manner ~\cite{ravanelli2020multi}. The SSL models can extract high-level, informative, and compact feature vectors from the raw audio inputs. 
The extracted features improve downstream tasks like speech recognition, speaker verification, spoken language understanding, etc.~\cite{yang21c_interspeech}. 
Only requiring unlabeled audios is a desirable property of SSL since large-scale unlabeled audios can be collected easily compared to labeled data. 

Nevertheless, the extreme size of the unlabeled corpus also makes it hard for the developers to ensure that there is not any private information in the corpus.
It is still possible that a malicious person can attack the SSL models to retrieve some sensitive information in the pre-training data.
In the natural language processing (NLP) community, researchers have successfully eavesdropped on sensitive information such as phone numbers from SSL NLP models~\cite{carlini2020extracting}. 
However, there is still a lack of systematic studies about the private information leakage of SSL speech models to our best knowledge.
As the self-supervised model becomes more and more ubiquitous, the practitioners may consider pre-training their models on some application-specific datasets which may contain sensitive information, such as the recording of the online clinic. 
Consider someone releasing an SSL speech model trained on online clinic recordings to benefit all online clinic applications worldwide. If MIA finds that the celebrity's voice involves in the pre-training dataset of the SSL model, it indicates that the celebrity used clinical services. However, medical records are very private and should not be disclosed to the public.
Consequently, it is imperative for us to either eliminate these concerns or verify the existence of the privacy leakage.

In this paper, we perform the first MIA against several SSL speech models under black-box access. 
The results show that SSL speech models are vulnerable to such attack at both speaker and utterance-level. 
Besides, we also conduct an ablation study to understand how the size of the model, the pre-trained dataset, and a simple data perturbation affect the attack performance.



\begin{figure*}[t]
    \centering
    \includegraphics[width=0.91\linewidth]{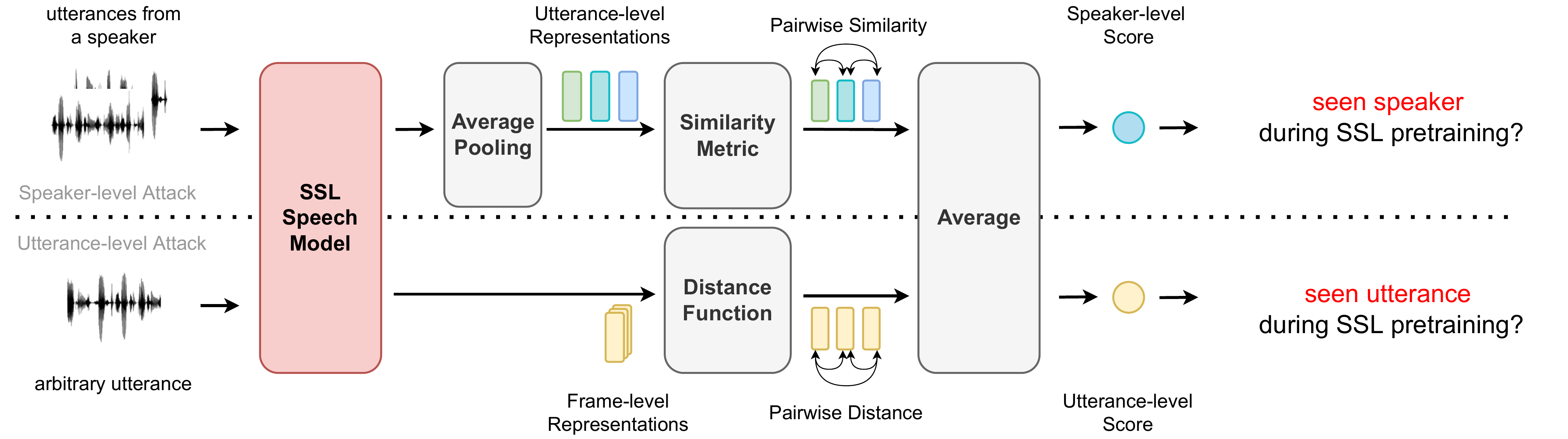}%
    
    \caption{The proposed membership inference attack against self-supervised pre-trained speech models.}
    \label{fig:attack}
\end{figure*}

\section{Methodology}
\label{sec:related-works}

\subsection{Threat Modeling}
\label{ssec:threat-model}
Here we consider an adversary who has black-box input-output access to the target SSL speech model $\mathcal{M}$ which has been pre-trained on dataset $D_{target}$ that possibly contains sensitive information. 
The adversary can only infer $\mathcal{M}$ with some utterances and compute their output representations while having no knowledge about both the structure and the pre-training algorithm of $\mathcal{M}$.
In addition, the adversary may utilize an auxiliary dataset $D_{aux}$ to perform MIA. 
$D_{aux}$ comprises aggregate utterances from several speakers that are not included in $D_{target}$.

Since SSL speech models are pre-trained on large-scale utterances from various speakers, the adversary could launch MIA at two levels, according to the speaker identities or the utterance information. We introduce the utterance-level attacks in section \ref{sec:utterance-attack} and the speaker-level attack in section \ref{sec:speaker-attack}. 
For the attack strategy, we adopt a similar approach to ~\cite{song2020information} to perform thresholding attack based on either similarity or distance measurements, in tandem with the observation that SSL speech models usually maximize certain similarity or distance between models' representations and some acoustic features in each utterance. For example, HuBERT~\cite{hsu2021hubert} maximizes the similarity between its representations and the centroids of the MFCC clusters; Wav2vec 2.0~\cite{baevski2020wav2vec} and CPC~\cite{oord2018representation} learn in a contrastive learning manner that minimizes the distance between the positive samples while keeping away from the negative samples. So we believe that these statistics of the data in $D_{target}$ and the unseen could be very different and distinguishable. The detailed methods are introduced in the next subsections.

\subsection{Utterance-level Attack}
\label{sec:utterance-attack}
In utterance-level attack, the adversary inputs a specific utterance to $\mathcal{M}$ and try to decide whether it belongs to the pre-training dataset of the target model $M$ or not.
The lower part of Figure \ref{fig:attack} illustrates the proposed utterance-level attack.
The attack falls into two stages, namely \emph{basic attack} and \emph{improved attack}. \emph{Basic attack} does not require any additional parameters and attacker models. While in \emph{improved attack}, we train a neural network to better distinguish the utterances.
\\
\textbf{Basic attack}:
in basic attack, we start with an utterance with $T$ sample points $\bm{x} = [x_1, x_2, \cdots, x_T]$. The adversary first encodes the utterance into a sequence of representations $\bm{H} = [\bm{h}_1, \bm{h}_2, \cdots, \bm{h}_m]$, where $m$ is the length of features of the utterance. Here $\bm{h}_i \in \mathbb{R}^{q}$
is referred to as frame-level representations, and $q$ is their dimensionality.
Then the adversary calculates the \textbf{utterance-level score} by averaging the distance between each frame-level representation:
\begin{equation*}
    S_{uttr} = \frac{2}{m(m-1)}\sum_{i=1}^m\sum_{j=i+1}^m d_{uttr}(\bm{h}_i, \bm{h}_j)
\end{equation*}
where $d_{uttr}: \mathbb{R}^{q} \times \mathbb{R}^{q} \mapsto \mathbb{R}$ is a distance measure function (e.g. euclidean distance, cosine distance).

Finally, the adversary uses $S_{uttr}$ to decide membership: if $S_{uttr}$ is above some pre-defined threshold, then the adversary will say $\bm{x}$ belongs to $D_{target}$; otherwise, it is not. 
\\
\textbf{Improved attack}:
using a pre-defined distance function may not achieve the best membership inference performance.
Here we adopt \textit{pseudo-labeling} on a subset of $D_{aux}$ to propose an improved version of utterance-level membership inference attack.
Given utterances $\{\bm{x}_i\}_{i=1}^N \in D_{aux}$, the adversary first compute $S_{uttr}$ for each $x_i$.
Then $k$ utterances with the highest $S_{uttr}$ are selected and pseudo-labeled as \emph{seen} data to the target model $M$, while the $k$ utterances with the lowest $S_{uttr}$ are also selected and pseudo-labeled as \emph{unseen}.
These selected utterances are then used to train a deep neural network  $f_{uttr}: \mathbb{R}^{m \times q} \mapsto \mathbb{R}$ that takes the frame-level representations of an utterance as input to predict the utterance-level score.
That is to say, in the improved version of utterance-level attack, the DNN is to replace the pre-defined distance function and the average operation of the basic attack. 
We use binary cross-entropy loss to train the deep neural network. 

\subsection{Speaker-level Attack}
\label{sec:speaker-attack}
In speaker-level attack, the adversary inputs aggregate utterances from a certain speaker to $\mathcal{M}$ and aims to determine whether this speaker involves in $D_{target}$ or not (which is the case describe in the end of Section \ref{sec:intro}). 
The upper part of Figure \ref{fig:attack} illustrates the proposed speaker-level attack.
Likewise, we divide the attack into two stages.
\\
\textbf{Basic attack}:
in basic attack, we start with the aggregation $\mathcal{X}_C$ of $n$ utterances $\{\bm{x}^C_i\}_{i=1}^n$ 
from certain speaker $C$. The adversary first computes the utterance-level representations $\{\bar{\bm{H}_i}\}_{i=1}^n$ by taking the average of the frame-level representations of each utterance $x_i^C$, as we believe that utterance-level representation may contain more speaker information. 
Then the adversary calculates the \textbf{speaker-level score} by averaging the similarity between the utterance-level representations:

\begin{equation*}
    S_{spkr} = \frac{2}{n(n-1)}\sum_{i=1}^n\sum_{j=i+1}^n \delta_{spkr}(\bar{\bm{H}_i}, \bar{\bm{H}_j})
\end{equation*}

where $\delta_{spkr}: \mathbb{R}^{q} \times \mathbb{R}^{q} \mapsto \mathbb{R}$ is a vector similarity metric function (e.g. cosine similarity).

Finally, the adversary uses $S_{spkr}$ to decide membership: If $S_{spkr}$ is above some pre-defined threshold, the adversary will say that $C$ involves in the $D_{target}$, and vice versa. 
\\
\textbf{Improved attack}:
for better performance on the speaker-level attack, we also utilize $D_{aux}$ to propose an improved version of speaker-level membership inference attack.
Given a set of aggregate utterances $\{\mathcal{X}_i\}_{i=1}^m$ from $m$ speakers in $D_{aux}$, the adversary first compute $S_{spkr}$ for each $\mathcal{X}_i$. 
Then the utterances of $k$ speakers with the highest $S_{spkr}$ are selected and pseudo-labeled as \emph{seen} data to the target model $M$, while the utterances of $k$ speakers with the lowest $S_{spkr}$ are also selected and labeled as \emph{unseen}.
These selected utterances are then used to train a deep neural network $f_{spkr}: (\mathbb{R}^{q}\times \mathbb{R}^{q}) \mapsto \mathbb{R}$ that takes a pair of utterance-level representations as input to predict the pairwise similarity. 
Hence, in the improved version of speaker-level attack, the DNN is to replace the pre-defined similarity metric of the basic attack.
We also use binary cross-entropy loss to train the network. 

\begin{figure*}[!t]
    \centering
    \begin{subfigure}[b]{0.225\linewidth}
        \includegraphics[width=\textwidth]{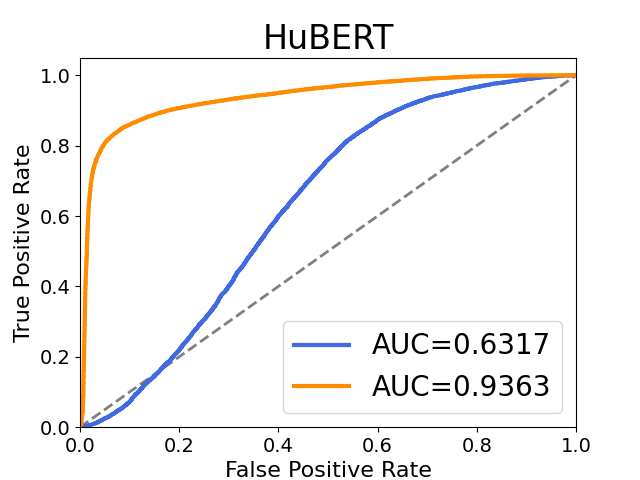}
    \end{subfigure}
    \hfill
    \begin{subfigure}[b]{0.225\linewidth}
        \includegraphics[width=\textwidth]{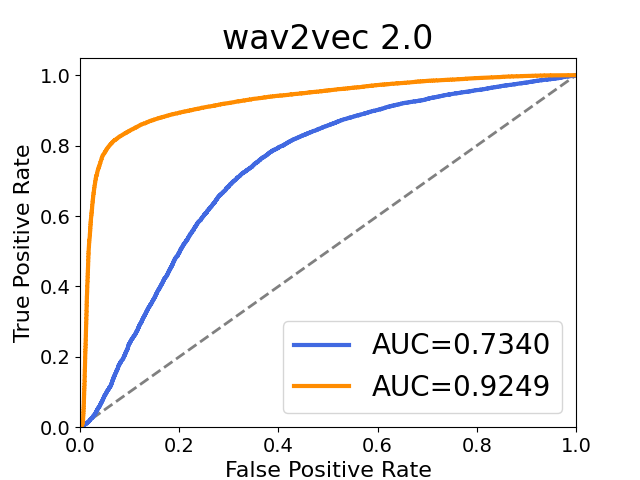}
    \end{subfigure}
    \hfill
    \begin{subfigure}[b]{0.225\linewidth}
        \includegraphics[width=\textwidth]{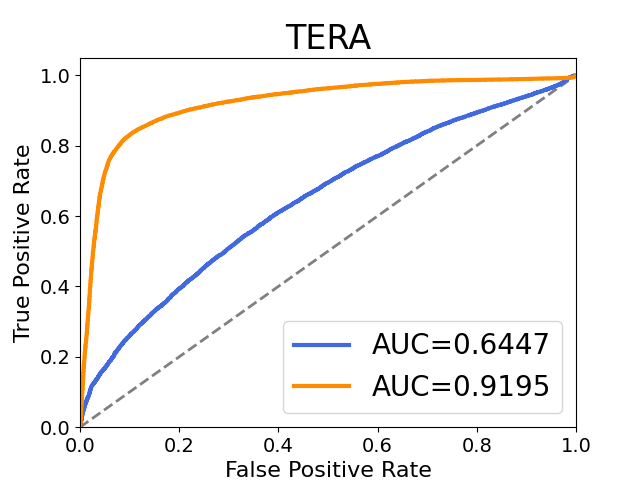}
    \end{subfigure}
    \hfill
    \begin{subfigure}[b]{0.225\linewidth}
        \includegraphics[width=\textwidth]{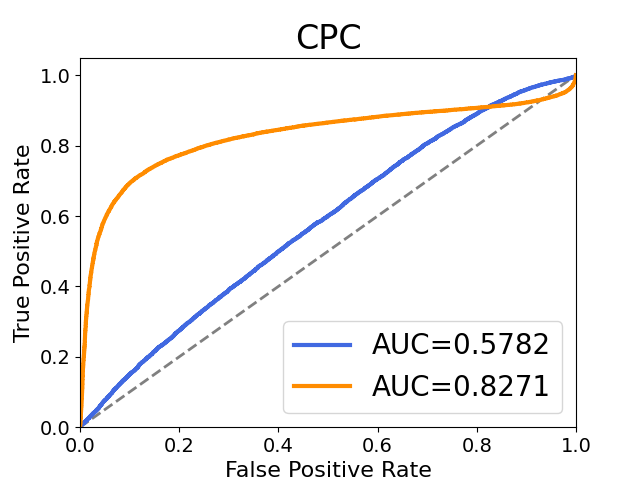}
    \end{subfigure}
    \caption{The ROC curve of the proposed utterance-level attack against four self-supervised speech models. The blue line is the ROC curve of basic attack while the orange line indicates the improved attack.}
\end{figure*}

\begin{figure*}[!t]
    \centering
    \begin{subfigure}[b]{0.225\linewidth}
        \includegraphics[width=\textwidth]{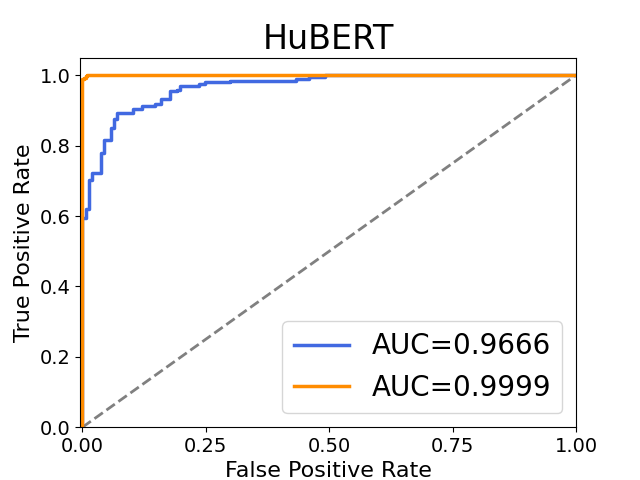}
    \end{subfigure}
    \hfill
    \begin{subfigure}[b]{0.225\linewidth}
        \includegraphics[width=\textwidth]{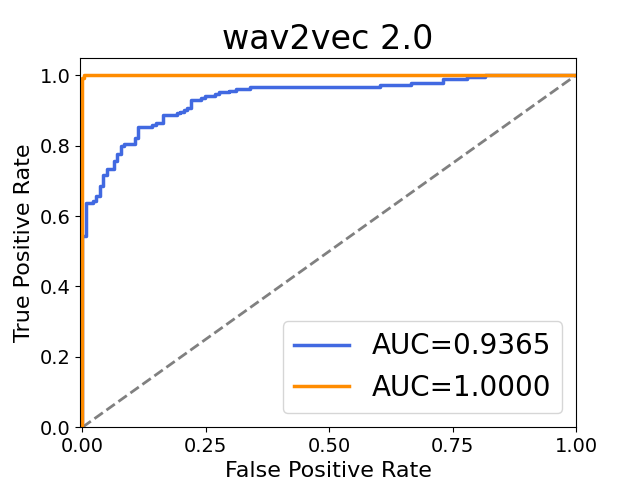}
    \end{subfigure}
    \hfill
    \begin{subfigure}[b]{0.225\linewidth}
        \includegraphics[width=\textwidth]{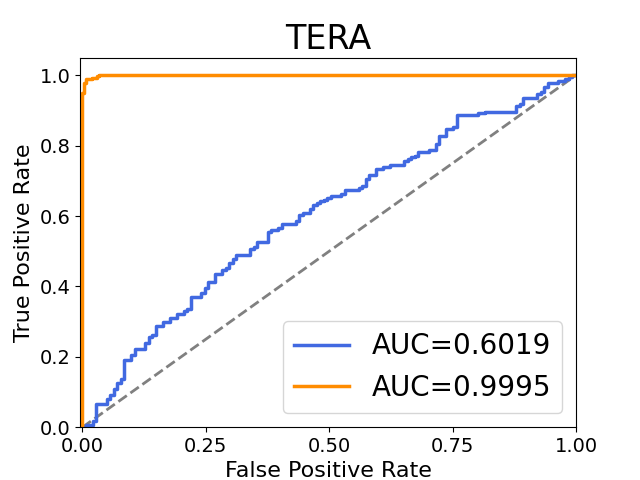}
    \end{subfigure}
    \hfill
    \begin{subfigure}[b]{0.225\linewidth}
        \includegraphics[width=\textwidth]{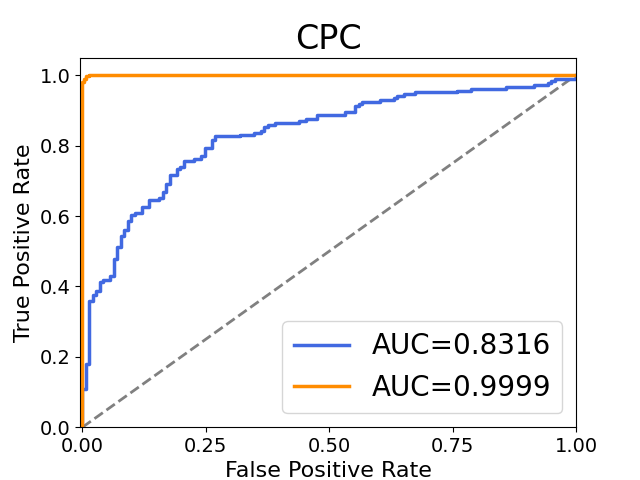}
    \end{subfigure}
    \caption{The ROC curve of the proposed speaker-level attack against four self-supervised speech models. The blue line is the ROC curve of basic attack while the orange line indicates the improved attack.}
\end{figure*}

\section{Experimental Setting}
In the experiments, we use four SSL speech models from S3PRL toolkits~\cite{S3PRL}: HuBERT, wav2vec 2.0, CPC, and TERA.
These models were pre-trained on large-scale unlabeled data such as LibriSpeech \cite{7178964} and Libri-Light \cite{kahn2020libri}. 
Five subsets of the LibriSpeech corpus are involved in the experiments: \emph{train-clean-100}, \emph{dev-clean}, \emph{dev-other}, \emph{test-clean} and \emph{test-other}, where \emph{train-clean-100} serves as seen for $\mathcal{M}$. The rest constitute $D_{aux}$ and serve as unseen. We also conduct experiments in which VCTK-Corpus~\cite{yamagishi2019vctk} servers as $D_{aux}$ to amplify the difference between $D_{target}$ and $D_{aux}$.

For utterance-level attack, motivated by the widely used contrastive loss \cite{chopra2005learning} in self-supervised learning, we use \emph{cosine distance} (defined as 1-\emph{cosine similarity}) for the predefined distance function $d_{uttr}$.
And for the DNN $f_{uttr}$ in the improved attack, it contains an attentive pooling layer \cite{safari2020self} followed by two linear layers. 
We optimize $f$ for 20 epochs with learning rate set to $10^{-5}$ and $k=500$.

As for the speaker-level attack, since we empirically find out that most speakers have similarity scores close to 1, so we use \emph{cosine similarity} for the predefined similarity metric $\delta_{spkr}$. 
And for the DNN $f_{spkr}$ in the improved attack, it contains an attention pooling layer followed by a linear layer and a dot product layer. 
We optimize $f$ for 20 epochs with learning rate set to $10^{-5}$ and $k=1$. 

    

\begin{table}[!h]
\centering
\caption{The Area Under the Curve (AUC) value when the unseen data comes from VCTK corpus.}
\begin{tabular}{l|c|c|c|c}
        \toprule
          & HuBERT & wav2vec 2.0 & TERA & CPC \\
         \midrule
         \multicolumn{5}{c}{$D_{aux}$ comes from LibriSpeech} \\
         \midrule
         utterance & 0.9363 & 0.9249 & 0.9195 & 0.8271 \\
         speaker & 0.9999 & 1.0000 & 0.9995 & 0.9999  \\
         \midrule
         \multicolumn{5}{c}{$D_{aux}$ comes from VCTK} \\
         \midrule
         utterance & 0.9380 & 0.9121 & 0.8757 & 0.9384 \\
         speaker & 0.9999 & 0.9979 & 0.9999 & 0.9998 \\
         \bottomrule
    \end{tabular}
\label{tab:vctk_ablation}
\end{table}

\section{Results}

\subsection{Results}
To evaluate the performance of membership inference attacks, a natural choice is to consider the tradeoff between the true-positive rate (TPR) and the false-positive rate (FPR).
Here we show the Receiver Operating Characteristic (ROC) curve of the proposed attack against SSL speech models at both utterance-level (Figure 1) and speaker-level (Figure 2).
Besides, we also report Area Under the Curve (AUC) for quantitative comparison.
Noted that the blue line represents the ROC curve of the basic attack while the orange line indicates the improved attack.

At the utterance-level attack, we can observe that all pre-trained models are quite vulnerable to such attack to reveal membership information except CPC. Wav2vec 2.0 even has AUC over 0.7 when we apply the basic attack.
Additionally, using DNN in the improved attack significantly boosts the attack performance, where the AUC of each pre-trained model is 0.9363 (HuBERT), 0.9249 (wav2vec 2.0), 0.9195 (TERA) and 0.8271 (CPC), respectively. 
And the proposed attack is even effective at a low FPR ($<$0.1).

Furthermore, at the speaker-level attack, the membership information leakage of most pre-trained models is more severe than the one in utterance-level except TERA. 
For the basic attack, the worst one, HuBERT, even has an AUC value of around 0.97.
Moreover, we surprisingly find out that with only $k=1$, using DNN can further improve the attack to obtain a near-optimal performance, where the AUC against all models is close to 1. 
The results remain similar if $D_{aux}$ comes from VCTK-Corpus, as shown in table \ref{tab:vctk_ablation}. So the success of the proposed attacks does not result from any special selections of $D_{aux}$.
These results indicate that SSL speech models are weak in preventing membership information leakage regardless of the self-supervised objectives. 
Especially, the state-of-the-art models, HuBERT and wav2vec 2.0, have the worst privacy-preserving ability, which validates the intuition -- the utility-privacy trade-off still exists when people apply SSL.


\subsection{Ablation Study}
\label{sec:exp-settings}

We then inspect how the number of parameters of the pre-trained models and the size of the pre-training dataset affects the attack performance by applying the attack against several variants of these models. 
For the number of parameters, we compare (1) HuBERT-\{Large, X-Large\} pre-trained on LibriLight-60Khr (LL-60K), and (2) wav2vec 2.0-\{Base, Large\} pre-trained on LibriSpeech-960hr (LS-960).
As for the size of the pre-training dataset, we compare wav2vec 2.0 models pre-trained on LS-960 or LL-60K, and TERA models pre-trained on LibriSpeech-100hr (LS-100) or LS-960. 
All of these models are published by their original authors to ensure the same pre-training policy. Here we only report the AUC value against each model under basic attack due to the space limitation.

The upper part of table \ref{tab:model_ablation} lists the AUC value of improved attack when the model size differs. We can observe that using a larger model leads to a lower attack performance in both utterance-level and speaker-level.
The lower part of table \ref{tab:model_ablation} shows the attack performance when the pre-training dataset size changes. We can see that in speaker-level MIA, pre-training the model on larger dataset results in a lower attack performance, preventing itself from membership information leakage.
But when it comes to utterance-level attack, there's no such guarantee which dataset size is better.
These results are surprising as most of the previous works demonstrated that using a larger model or utilizing more training data may reduce the hassle of privacy risks \cite{carlini2020extracting, melis2019exploiting, shokri2021privacy}.
This strongly motivates the need for an in-depth study on the influence of these factors and the development of privacy-preserving techniques for pre-training SSL speech models in the future.

\begin{table}[t]
\centering
\caption{The Area Under the Curve (AUC) of the proposed attack against SSL model variants when either the model size (upper part) or dataset size (lower part) varies.}
\begin{subtable}[h]{0.98\linewidth}
    \centering
    \begin{tabular}{l|c|c|c|c}
        \toprule
         \multirow{2}{*}{Model} & \multicolumn{2}{c|}{HuBERT} & \multicolumn{2}{c}{wav2vec 2.0} \\
         \cmidrule{2-5}
         {} & Large & X-Large & Base & Large \\
         \midrule
         utterance& \textbf{0.7746} & 0.7487 & \textbf{0.6672} & 0.6342 \\
         speaker & \textbf{0.9711} & 0.9103 & \textbf{0.9300} & 0.8691 \\
         \bottomrule
    \end{tabular}
\end{subtable}
\par\bigskip 
\begin{subtable}[h]{0.98\linewidth}
    \centering
    \begin{tabular}{l|c|c|c|c}
         \toprule
         \multirow{2}{*}{Model} & \multicolumn{2}{c|}{wav2vec 2.0} & \multicolumn{2}{c}{TERA} \\
         \cmidrule{2-5}
         {}  & LS-960 & LL-60K & LS-100 & LS-960 \\
         \midrule
         utterance & \textbf{0.7340} & 0.6672 & 0.5835 & \textbf{0.6447} \\
         speaker & \textbf{0.9365} & 0.9300 & \textbf{0.7744} & 0.6019 \\
         \bottomrule
    \end{tabular}
\end{subtable}
\label{tab:model_ablation}
\end{table}


\section{Bypass simple defense}
\label{sec:defense}

An intuitive way to alleviate from the proposed attack may be pre-training SSL models with strong data augmentations. 
If attackers do not properly augment their auxiliary datasets, the proposed attack could possibly fail and always return "unseen". 
In this section, we would like to demonstrate that the proposed attack could potentially bypass this kind of defense. 
We start with pre-training a TERA model on LS-100 with all waveforms reversed (called TERA-reverse) because a normal waveform and a reversed one sound very different for humans. 
On several tasks such as phoneme classification and speaker verification, the performance of this model is close to the one pre-trained on normal LS-100. 
We then perform improved attack against it with normal waveforms, as shown in table \ref{tab:reverse_tera}.
Reversing waveforms negligibly reduces utterance-level privacy leakage and conversely deteriorates speaker-level leakage.

We further extend our experiment to other SSL models. 
Due to resource limitations, however, we could not fully pre-train these models with the reversed LS-960 or LL-60K from scratch. 
So we consider an approximated scenario and try to get some insights. 
In the scenario, the SSL models are pre-trained on normal waveforms while the attackers perform MIA with the reversed waveforms. 
We speculate that its effect is close to pre-training on reversed waveforms and attacking the models with normal waveforms. 
The results are shown in table \ref{tab:reverse_ablation}.
Attacking with reversed waveform decreases the performance of the improved utterance-level attack on CPC and TERA but has marginal effects on more advanced models, HuBERT and Wav2Vec 2.0. 
Consistent with the observation in table \ref{tab:reverse_tera}, using reversed waveforms is harmful for protecting speaker-level information.
Overall, the privacy issues of SSL speech models are critical. 
Only reversing waveform may not be enough to resolve them. 
Especially, more advanced techniques are required if we would like to keep the speaker information secure.
We leave this issue and the combination of other defense methods and SSL speech models for the future work.

\begin{table}[t]
\centering
\caption{The Area Under the Curve (AUC) of the proposed attack against TERA models pre-trained on normal LS-100 and the reversed one (TERA-reverse).}
\begin{tabular}{l|c|c}
        \toprule
          & TERA (LS-100) & TERA-reverse \\
         \midrule
         utterance & 0.8431 & 0.7645  \\
         speaker & 0.9952 & 0.9997 \\
         \bottomrule
    \end{tabular}
\label{tab:reverse_tera}
\end{table}

\begin{table}[!t]
\centering
\caption{The Area Under the Curve of the proposed attack when the attacker applies MIA with the reversed waveforms.}
\begin{tabular}{l|c|c|c|c}
        \toprule
          & HuBERT & wav2vec 2.0 & TERA & CPC \\
         \midrule
         \multicolumn{5}{c}{Attack with normal waveforms} \\
         \midrule
         utterance & 0.9363 & 0.9249 & 0.9195 & 0.8271 \\
         speaker & 0.9999 & 1.0000 & 0.9995 & 0.9999  \\
         \midrule
         \multicolumn{5}{c}{Attack with reversed waveforms} \\
         \midrule
         utterance & 0.9316 & 0.9269 & \textbf{0.6155} & \textbf{0.8003}\\
         speaker & 0.9998 & 1.0000 & 0.9995 & 1.000  \\
         \bottomrule
    \end{tabular}
\label{tab:reverse_ablation}
\end{table}

\section{Conclusions}
\label{sec:conclusions}

This paper performs the first membership inference attack against self-supervised pre-trained speech models under black-box access. 
The results show that these models are vulnerable to both speaker and utterance-level attacks.
We also conduct an ablation study indicating that with smaller datasets, one can slightly reduce the risk of privacy leakage, which is different to the observation of previous works.
The success of the proposed attacks suggests that the representations of SSL models encode the membership information of the pre-training data, which can cause severe privacy issues. 
We also conduct a preliminary study of defense but find that a simple data augmentation is not enough to prevent the proposed attack.
This strongly gives rise to the need for caution and motivates demands for developing privacy-preserving pre-training techniques in the future.

\section{Acknowledgement}

We thank to National Center for High-performance Computing (NCHC) of National Applied Research Laboratories (NARLabs) in Taiwan for providing computational and storage resources.

\bibliographystyle{IEEEtran}

\bibliography{mybib}


\end{document}